\documentclass{elsart}

\usepackage[english]{babel}
\usepackage{amsmath,amssymb}
\usepackage{graphicx}

\newcommand{\EtAl}{{\it et al.}}

\newcommand{\eff}{\mathrm{eff}}
\newcommand{\eps}{\varepsilon}
\newcommand{\tht}{\vartheta}

\begin{document}

\raisebox{8mm}[0pt][0pt]{\hspace{12cm}\vbox{hep-ph/0112310\\IFIC/01-71}}

\begin{frontmatter}

\title{Status of a hybrid three-neutrino interpretation of neutrino data}

\author[UNICAMP]{M.~M.\ Guzzo},
\ead{guzzo@ifi.unicamp.br}
\author[ICTP]{P.~C.\ de Holanda},
\ead{holanda@ictp.trieste.it}
\author[IFIC]{M.\ Maltoni},
\ead{maltoni@ific.uv.es}
\author[UNICAMP,UNISAO]{H.\ Nunokawa},
\ead{nunokawa@ift.unesp.br}
\author[IFIC]{M.~A.\ T{\'o}rtola} and
\ead{mariam@ific.uv.es}
\author[IFIC]{J.~W.~F.\ Valle}
\ead{valle@ific.uv.es}

\address[UNICAMP]{Instituto de F\'{\i}sica Gleb Wataghin,
  Universidade Estadual de Campinas - UNICAMP
  P.O. Box 6165, 13083-970 Campinas SP Brazil}

\address[ICTP]{ICTP, Strada Costiera 11, Miramare - Grignano,
  I-34014 Trieste, Italy}

\address[IFIC]{Instituto de F\'{\i}sica Corpuscular -- C.S.I.C.,
  Universitat de Val{\`e}ncia \\
  Edificio Institutos de Paterna, Apt.\ 22085, E--46071 Val{\`e}ncia, Spain}

\address[UNISAO]{Instituto de F\'{\i}sica Te{\'o}rica,
  Universidade Estadual Paulista, Rua Pamplona 145 \\
  01405-900 S{\~a}o Paulo, SP Brazil }

\begin{abstract}
    We re-analyse the non-standard interaction (NSI) solutions to the
    solar neutrino problem in the light of the latest solar as well as
    atmospheric neutrino data. The latter require oscillations (OSC),
    while the former do not. Within such a three-neutrino framework
    the solar and atmospheric neutrino sectors are connected not only
    by the neutrino mixing angle $\tht_{13}$ constrained by reactor
    and atmospheric data, but also by the flavour-changing (FC) and
    non-universal (NU) parameters accounting for the solar data.
    Since the NSI solution is energy-independent the spectrum is
    undistorted, so that the global analysis observables are the solar
    neutrino rates in all experiments as well as the Super-Kamiokande
    day-night measurements.  We find that the NSI description of solar
    data is slightly better than that of the OSC solution and that the
    allowed NSI regions are determined mainly by the rate analysis.
    By using a few simplified {\it ans\"atze} for the NSI interactions
    we explicitly demonstrate that the NSI values indicated by the
    solar data analysis are fully acceptable also for the atmospheric
    data.
    
    In the appendix we present an updated analysis combining the
    latest data from all solar neutrino experiments with the first
    results from KamLAND. We show that, although NSI still gives an
    excellent description of the solar data, the inclusion of KamLAND
    excludes at more than 3$\sigma$ the NSI hypothesis as a solution
    to the solar neutrino problem.
    
    \begin{keyword}
        neutrino oscillations \sep solar and atmospheric neutrinos
        \sep neutrino mass and mixing \sep non-standard neutrino
        interactions
        \PACS 14.60.Pq \sep 14.60.Lm \sep 26.65.+t \sep 13.15.+g
    \end{keyword}
\end{abstract}

\end{frontmatter}

\section{Introduction}
\label{sec:introduction}

The wealth of data from solar~\cite{sun-exp,Ahmad:2001an} and
atmospheric neutrinos~\cite{skatm,atm-exp,macroOld,macroNew} has put
neutrino physics in the spotlight and physicists are asking what is it
that makes solar and atmospheric neutrinos convert. The most popular
possibility is that of neutrino oscillations, expected in most
theories of neutrino mass~\cite{theory}.  Indeed an excellent joint
description of solar and atmospheric data is obtained in this
case~\cite{Gonzalez-Garcia:2001sq,others}. However, more often than
not, models of neutrino masses are accompanied by non-standard
interactions (NSI) of neutrinos~\cite{theory+nsirev}. These can have
non-universal (NU) or flavour-changing (FC) components, which
typically co-exist. The simplest NSI does not involve additional
interactions beyond those mediated by the Standard Model electroweak
gauge bosons: it is simply the nature of the leptonic charged and
neutral current interactions which is non-standard because of the
complexity of neutrino mixing~\cite{theory}. On the other hand NSI can
also be mediated by the exchange of new particles with mass at the
weak scale such as in some supersymmetric models with
R-parity-violating~\cite{Ross:1985yg,Hall:1984id} interactions.
Such non-standard flavour-violating physics can arise even in the
absence of neutrino mass~\cite{NSImodels2,NSImodels3}.

Such varieties of non-standard interactions of neutrinos affect their
propagation in matter~\cite{MSW}. The magnitude of the effect depends
on the interplay between conventional mass-induced neutrino
oscillation features in matter and those genuinely engendered by the
NSI, which do not require neutrino
mass~\cite{Valle:1987gv,NSIrecent0}.
These may have a variety of phenomenological
implications~\cite{NSIrecent0,NSIrecent1} and have been considered in
the context of atmospheric~\cite{Gonzalez-Garcia:1999hj,Val}, as well
as astrophysical neutrino sources~\cite{NSIastro}. The impact of
non-standard interactions of neutrinos has also been considered from
the point of view of future experiments involving solar
neutrinos~\cite{Berezhiani:2001rt} as well as the upcoming neutrino
factories~\cite{Huber:2001de,Huber:2001zw}.

In this paper we re-consider the case of NSI solutions to the solar
neutrino problem~\cite{Gago:2001si} taking into account the recent
charged current measurement at the Sudbury Neutrino Observatory
(SNO)~\cite{Ahmad:2001an}, the 1258--day Super-Kamiokande (SK) data
and the previous solar neutrino data~\cite{sun-exp}.  In contrast to
previous work we consider a three-neutrino NSI analysis and study also
the dependence of the solar neutrino NSI solution on the neutrino
mixing angle $\tht_{13}$ whose value is presently constrained mainly
by the reactor neutrino data~\cite{Apollonio:1999ae}.  Due to its
energy-independent nature, the NSI solution is in excellent agreement
with the flat spectral energy distribution observed in the
Super-Kamiokande experiment. Thus we focus first on the determination
of the allowed solutions by considering only the total rates of the
solar neutrino experiments. We find that these solutions provide
excellent descriptions of the solar rates, including the recent SNO
charged current (CC) result. Then we study the impact of the day-night
data from Super-Kamiokande measurements and show that the NSI
solutions are consistent with the non-observation of day-night
variations. Finally, and more important, using simplified {\it
  ans\"atze} for the NSI interactions we analyse the impact of the NSI
description of solar data on the atmospheric data, showing how they
are fully acceptable also for the latter.

This paper is organized as follows. In section
\ref{sec:neutr-evol-pres} we discuss the neutrino evolution and
conversion probabilities in the presence of NSI, in section
\ref{sec:analysis-method} we summarize the calculational and fit
procedures we adopt. In section \ref{sec:how-about-atmosph} we discuss
the impact of the NSI solar solution on the atmospheric data analysis,
and in section \ref{sec:conclusions} we summarize our results.

\section{Neutrino Evolution in the presence of NSI}
\label{sec:neutr-evol-pres}

The most general form of three-neutrino evolution Hamiltonian in the
flavor base $(\nu_e, \nu_\mu, \nu_\tau)$, in the presence of NSI can
be given as
\begin{equation}
    \mathbf{H} = \frac{1}{2E} \mathbf{R}
    \begin{pmatrix}
        0 & 0 & 0 \\
        0 & 0 & 0 \\
        0 & 0 & \Delta m_{32}^2
    \end{pmatrix}
    \mathbf{R}^\dagger + \sqrt{2} G_F N_e(\vec{r})
    \begin{pmatrix}
        1 & 0 & 0 \\
        0 & 0 & 0 \\
        0 & 0 & 0
    \end{pmatrix}
    + \sqrt{2} G_F N_f(\vec{r})
    \begin{pmatrix}
        0         & \eps_{12} & \eps_{13} \\
        \eps_{12} & \eps_{22} & \eps_{23} \\
        \eps_{13} & \eps_{23} & \eps_{33}
    \end{pmatrix}
    \,;
    \label{eq:1}
\end{equation}
where E is the energy, $G_F$ Fermi's constant, $\Delta m^2_{32} \equiv
m^2_3 - m^2_2 \equiv \Delta m^2_{\rm atm}$ and $N_e(\vec{r})$ is the
number density of electrons along the neutrino trajectory.  In our
calculations of the $\nu_e$ survival probability we use the electron
and neutron number densities in the Sun from the SSM~\cite{SSM}, while
for the Earth matter effects we use the density profile given in the
Preliminary Reference Earth Model (PREM)~\cite{PREM}.

In Eq.~\eqref{eq:1} $\eps_{ij}$ parametrize the deviation from
standard neutrino interactions. For example $\sqrt{2} \, G_F
N_f(\vec{r}) \eps_{1\ell}$ is the forward scattering amplitude of the
FC process $\nu_e + f \to \nu_\ell + f$, while $\sqrt{2} \, G_F
N_f(\vec{r}) \eps_{\ell\ell}$ represents the difference between the
non-standard component of the $\nu_e + f$ and the $\nu_\ell + f$
elastic forward scattering amplitudes. The quantity $N_f(\vec{r})$ is
the number density of the fermion $f$ along the path $\vec{r}$ of the
neutrinos propagating in the Sun or the Earth.  We consider two cases,
depending on the NSI model, in which the neutrino interaction occurs
with down--type or up--type quarks. For the case of non-standard
interactions on electrons~\cite{Berezhiani:2001rt} one would have also
to take into account the effect of the NSI also in the neutrino
detection cross section.

Note that for simplicity we have set the mass splitting $\Delta
m_{21}^2$ of the first two neutrinos to zero, so that the
corresponding mixing angle $\tht_{12}$ can be eliminated.  This is
justified as we are mainly interested in isolating the effect of the
NSI in the description of solar neutrino data.  In contrast we assume
the most general NSI flavour structure, consistent with CP
conservation.  Under these assumptions, we
have~\cite{Schechter:1980bn}:
\begin{equation}
    \mathbf{R} =
    \begin{pmatrix}
        c_{13}          & 0       & s_{13}        \\
        - s_{23} s_{13} &  c_{23} & s_{23} c_{13} \\
        - s_{13} c_{23} & -s_{23} & c_{23} c_{13}
    \end{pmatrix},
\end{equation}
where we used the notation $c_{ij} \equiv \cos\tht_{ij}$ and $s_{ij}
\equiv \sin\tht_{ij}$. Similarly to the usual oscillation
case~\cite{2nueff} the $\nu_e$ survival probability $P_{ee}$ can be
written as
\begin{equation}
    P_{ee} = c_{13}^4 P_{ee}^\eff + s_{13}^4,
\end{equation}
where $P_{ee}^\eff$ is the electron survival probability in an
effective $2 \times 2$ model described by the Hamiltonian
\begin{equation}
    \mathbf{H}^\eff \equiv
    \left[
        \mathbf{R}^\dagger \mathbf{V} \mathbf{R}
    \right]_{2\times 2}
    = \sqrt{2} G_F N_e(\vec{r})
    \begin{pmatrix}
        c_{13}^2 & 0 \\
        0        & 0
    \end{pmatrix}
    + \sqrt{2} G_F N_f(\vec{r})
    \begin{pmatrix}
        0         & \eps_\eff \\
        \eps_\eff & \eps'_\eff
    \end{pmatrix}
\end{equation}
with the effective $\eps_\eff$ and $\eps'_\eff$ given in terms of the
original $\eps_{ij}$ parameters as
\begin{align}
    \label{eq:4} \eps_\eff
    &= c_{13} ( \eps_{12} c_{23} - \eps_{13} s_{23} )
    - s_{13} [ \eps_{23} (c_{23}^2 - s_{23}^2)
    + (\eps_{22} - \eps_{33}) c_{23} s_{23}] \,,
    \\
    \begin{split}
        \label{eq:5} \eps'_\eff
        &= \eps_{22} c_{23}^2 - 2 \eps_{23} c_{23} s_{23}
        + \eps_{33} s_{23}^2
        + 2 s_{13} c_{13} (\eps_{13} c_{23} + \eps_{12} s_{23}) \\
        &\quad
        - s_{13}^2 ( \eps_{33} c_{23}^2 + \eps_{22} s_{23}^2
        + 2 \eps_{23} s_{23} c_{23} ) \,.
    \end{split}
\end{align}
Note that even though we have assumed the most general NSI flavour
structure, the final propagation of solar neutrinos can be described
effectively as a two-dimensional evolution Hamiltonian depending only
on the mixing angle $\tht_{13}$ and on two effective NSI parameters
$(\eps_\eff, \eps'_\eff)$.

\section{Analysis Method} \label{sec:analysis-method}

In our description of the solar neutrino data~\cite{sun-exp} we adopt
the same analysis techniques already described in
Refs.~\cite{Gonzalez-Garcia:2001sq,two,Bahcall:2001zu} using the
latest theoretical standard solar model (SSM) best--fit fluxes and
estimated uncertainties~\cite{SSM}.  For the neutrino conversion
probabilities we use the results calculated numerically as indicated
above.

Since the parameter space is three-dimensional, the allowed regions
for a given C.L.\ are defined as the set of points satisfying the
condition
\begin{equation}
    \chi^2_{\rm sol}(\eps_\eff, \eps'_\eff, \tht_{13}  )
    -\chi^2_{\rm sol,min}\leq \Delta\chi^2 \mbox{(C.L., 3~d.o.f.)} ,
    \label{eq:chirsf}
\end{equation}

In our numerical calculations we use the survival/conversion
probabilities of solar electron neutrinos valid over a generous range
of NSI parameters $(\eps_\eff, \eps'_\eff)$. On the other hand, we use
the relevant reaction cross sections and efficiencies for the all
experiments employed in
Ref.~\cite{Gonzalez-Garcia:2001sq,two,Bahcall:2001zu}. For the SNO
case the CC cross section for deuterium was taken from~\cite{crossno}.

For what concerns the fit of the total rates, we take into account the
results of the Homestake, GALLEX/GNO, SAGE and Super-Kamiokande
experiments, together with the latest SNO CC result.  Thus we have $5$
experimental data, which we fit in terms of the $3$ parameters
$\eps_\eff$, $\eps'_\eff$ and $\tht_{13}$; therefore, the total number
of degrees of freedom is $2$.

In addition to total rates, we also take into account the
Super-Kamiokande ``zenith angle spectrum''~\cite{sun-exp}, which
includes both the spectral information on the final lepton energy (8
bins) and the information on the zenith angle distribution which
results from neutrino interactions inside the Earth (1-day and 6-night
bins).  This data sample contains a total of $44=6\times 7 +2$ bins.
Thus the number of degrees of freedom is 5 (rates) + 44
(zenith-spectrum) - 1 (free normalization) - 3 (fit parameters) = 45.

\begin{figure}[t] \centering
    \includegraphics[width=0.85\textwidth]{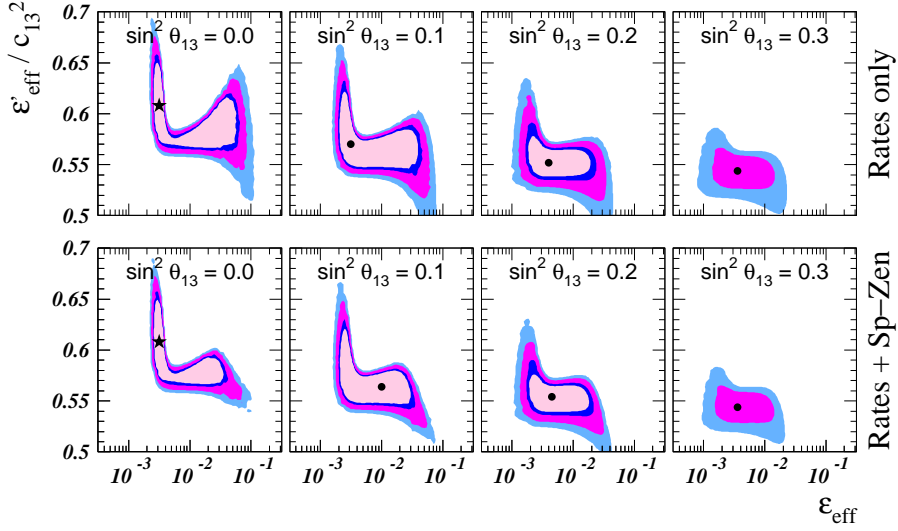}
    \caption{\label{fig:quark-D} %
      Allowed NSI regions indicated by the fit of the solar rates
      (upper panels) and global solar data (lower panels) for
      different values of $\tht_{13}$, assuming non-standard
      interactions of neutrinos with $d$-type quarks.  The shaded
      areas refer to the 90\%, 95\%, 99\% and 99.7\% C.L.\ with 3
      degrees of freedom. The best fit point is indicated by a star.}
\end{figure}
\begin{figure}[t] \centering
    \includegraphics[width=0.85\textwidth]{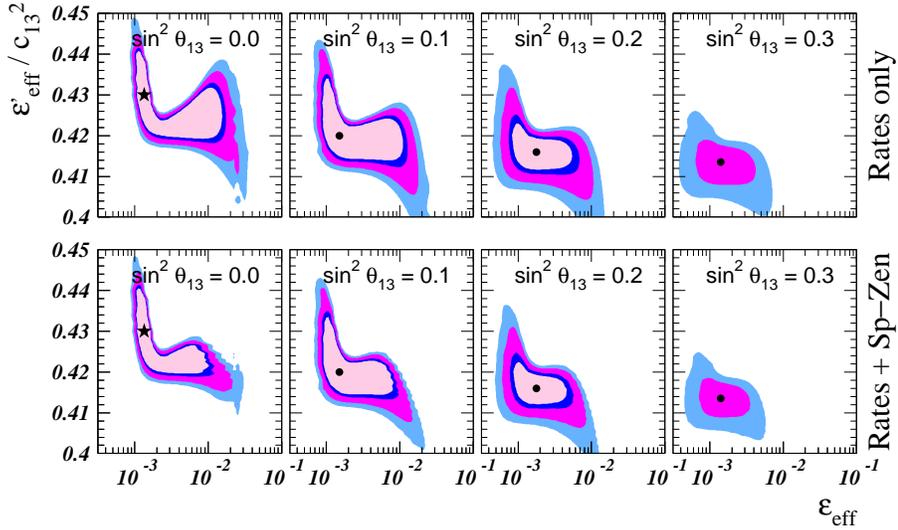}
    \caption{\label{fig:quark-U} %
      Same as Fig.~\ref{fig:quark-D} but for the case of non-standard
      interactions of neutrinos with $u$-type quarks.}
\end{figure}

In Figs.~\ref{fig:quark-D} and \ref{fig:quark-U} we present the
allowed regions in the $(\eps_\eff, \eps'_\eff)$ parameter space for
different values of $\sin^2\tht_{13}$, assuming non-standard
interactions of neutrinos with $d$-type (Fig.~\ref{fig:quark-D}) and
$u$-type (Fig.~\ref{fig:quark-U}) quarks. The shaded areas refer to
the 90\%, 95\%, 99\% and 99.7\% C.L.\ with 3 degrees of freedom, and
the best fit point is indicated by a star.
Let us now comment some of their features.  The first thing to notice
from Figs.~\ref{fig:quark-D} and \ref{fig:quark-U} is that the amount
of NU is large. Nevertheless this is allowed by experiment and
therefore consistent. We note also that the NSI strength required in
the case of interaction with u-type quarks is smaller than for the
case of d-type quarks.

\begin{figure}[t] \centering
    \includegraphics[width=0.7\textwidth]{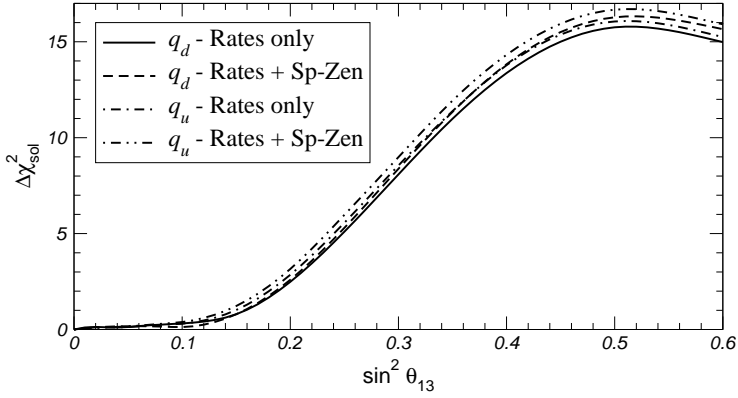}
    \caption{\label{fig:chisq} %
      Dependence of $\Delta \chi^2$ as a function of $\sin^2
      \tht_{13}$, for different NSI solutions.}
\end{figure}

Notice also that $\eps_\eff$ can only be large over a very narrow
$\eps'_\eff$ region close to 0.57, a fact which will be used shortly.
More importantly, we note that the quality of the solar fit becomes
progressively worse as $\tht_{13}$ increases, in a way similar to what
happens in a three-neutrino OSC
scenario~\cite{Gonzalez-Garcia:2001sq,teta13}. This indicates that,
although with less sensistivity, $\tht_{13}$ can be constrained solely
by solar data, irrespective of reactor data, also in the context of
the NSI mechanism.
In Fig.~\ref{fig:chisq} we present the dependence of $\Delta \chi^2$
as a function of $\sin^2 \tht_{13}$.
The {\it solid} line refers to non-standard interactions of neutrinos
with $d$-type quarks, and only the information from total rates is
used, while for the {\it dashed} line also the spectrum-zenith
distribution of Super-Kamiokande is included in the fit. The two {\it
  dotted-dashed} lines refer to non-standard interactions of neutrinos
with $u$-type quarks, both with and without the spectrum-zenith
information.

Finally in Table~\ref{tab:rates} we present the best-fit points and
goodness-of-fit of oscillation and NSI solutions to the solar neutrino
problem. Since the neutrino mixing angle $\tht_{13}$ is strongly
constrained by reactor neutrino data~\cite{Apollonio:1999ae} we set,
for definiteness, $\tht_{13}=0$ in what follows. This way we have
only two degrees of freedom.
\begin{table}[b] \centering
    \catcode`?=\active \def?{\hphantom{0}}
    \newcommand{\E}[2]{{#1}\times 10^{#2}}
    \begin{tabular}{lcccccc}
        \hline
        \noalign{\smallskip}
        Solution & $\Delta m_{21}^2$ & $\tan^2(\tht_{12})$ &
        $\eps_\eff$ & $\eps'_\eff$ &
        $\chi^2_\mathrm{min}$ & G.O.F. \\
        \noalign{\smallskip}
        \hline
        \noalign{\smallskip}
        LMA & $\E{2.4}{-5}$ & $0.31$        & $0$           & $0$    & $2.2?$ & $53\%$ \\
        SMA & $\E{7.8}{-6}$ & $\E{1.6}{-3}$ & $0$           & $0$    & $4.0?$ & $26\%$ \\
        NSI (d) & $0$       & $0$           & $\E{3.2}{-3}$ & $0.61$ & $0.60$ & $90\%$ \\
        NSI (u) & $0$       & $0$           & $\E{1.3}{-3}$ & $0.43$ & $0.62$ & $89\%$ \\
        \noalign{\smallskip}
        \hline
    \end{tabular}
    \caption{\label{tab:rates}%
      Best-fit points and goodness-of-fit of oscillation and NSI
      solutions to the solar neutrino problem, including only rates.}
\end{table}
The numbers in the table refer to a restricted analysis using only
$\Delta m_{21}^2$ and the neutrino mixing angle $\tht_{12}$ for the
pure OSC case and only $\eps_\eff$ and $\eps'_\eff$ for the pure NSI
case. We see that the pure two-parameter NSI solution is somewhat
better than the corresponding pure OSC solution.

\section{How about atmospheric neutrinos?}
\label{sec:how-about-atmosph}

So far we have given a description of solar neutrino data in terms of
NSI interactions.  We now turn to the issue of atmospheric neutrinos.
In principle one could imagine a pure NSI description of the
atmospheric data itself~\cite{Gonzalez-Garcia:1999hj}. However, it has
recently been shown that, while it may fit the contained atmospheric
data, such a description can not reconcile them with the up-going muon
data.
The sensitivity of the atmospheric data to the NSI follows mainly from
the fact that the wide energy range of the atmospheric neutrino data
sample, from the sub-GeV events up to the up-going samples of MACRO
and Super-Kamiokande~\cite{atm-exp,skatm,macroOld,macroNew} is enough
to manifest the energy-dependence characteristic of the atmospheric
neutrino conversion mechanism.
On this basis it has been shown how, taken altogether, the atmospheric
data leave very little room for NSI and can tolerate the existence of
NSI only at a sub-leading level~\cite{Val}.

The main question for us then is to evaluate whether the NSI
description of solar neutrino data is in conflict with the atmospheric
data sample. The danger lies in the fact that, in contrast to a pure
oscillation description of the neutrino
data~\cite{Gonzalez-Garcia:2001sq}, even when the value of the third
neutrino mixing angle $\tht_{13}$ (presently constrained mainly by the
reactor neutrino data) is taken to zero, the solar and atmospheric
sectors are still coupled with each other through the postulated NSI
interactions whose strength is fixed, as described above, in order to
account for the solar data.

A full description of the atmospheric and solar neutrino data in terms
of a hybrid OSC+NSI description is, at the moment, prohibitive in view
of the complexity of the problem. It suffices to remind the reader
that we expect ten relevant independent parameters in this case: in
addition to the five relevant CP conserving oscillation parameters of
Ref.~\cite{Gonzalez-Garcia:2001sq} (two mass splittings and three
angles) there are five new NSI parameters characterizing the magnitude
of FC and NU CP conserving non-standard interactions.

Thus we are forced to make a simplifying {\it ansatz}.  For
definiteness we consider the two following choices of the NSI matrix:
\begin{equation} \label{eq:ans_A}
    (a): \quad
    \begin{pmatrix}
        0         & \eps_{12} & \eps_{13} \\
        \eps_{12} & \eps_{22} & \eps_{23} \\
        \eps_{13} & \eps_{23} & \eps_{33}
    \end{pmatrix}
    =
    \begin{pmatrix}
        0              & \eps/\sqrt{2} & -\eps/\sqrt{2} \\
        \eps/\sqrt{2}  & \eps'         & 0              \\
        -\eps/\sqrt{2} & 0             & \eps'
    \end{pmatrix}
\end{equation}
and
\begin{equation} \label{eq:ans_B}
    (b): \quad
    \begin{pmatrix}
        0         & \eps_{12} & \eps_{13} \\
        \eps_{12} & \eps_{22} & \eps_{23} \\
        \eps_{13} & \eps_{23} & \eps_{33}
    \end{pmatrix}
    =
    \begin{pmatrix}
        0               & 0             & -\sqrt{2}\,\eps \\
        0               & \eps'         & 0               \\
        -\sqrt{2}\,\eps & 0             & \eps'
    \end{pmatrix}.
\end{equation}
With each of these choices the five new NSI parameters are reduced to
the two parameters involved in the description of the solar data, as
explained previously in sections \ref{sec:neutr-evol-pres} and
\ref{sec:analysis-method}. In fact for both of these choices, once we
set $\tht_{13}=0$ and $\tht_{23}=45^\circ$ (see below), we have
simply $\eps_\eff = \eps$ and $\eps_\eff' = \eps'$. For definiteness
we focus here on the case of neutrino non-standard interactions with
down-type quarks in Eq.~\eqref{eq:1}.

As for the data here we use the totality of sub- and multi-GeV ($e$,
$\mu$) atmospheric neutrino data~\cite{skatm,atm-exp} as well as
Super-Kamiokande stop and through-going muon data, and MACRO
through-going muons~\cite{macroOld,macroNew}.
\begin{figure}[t] \centering
    \includegraphics[width=0.99\textwidth]{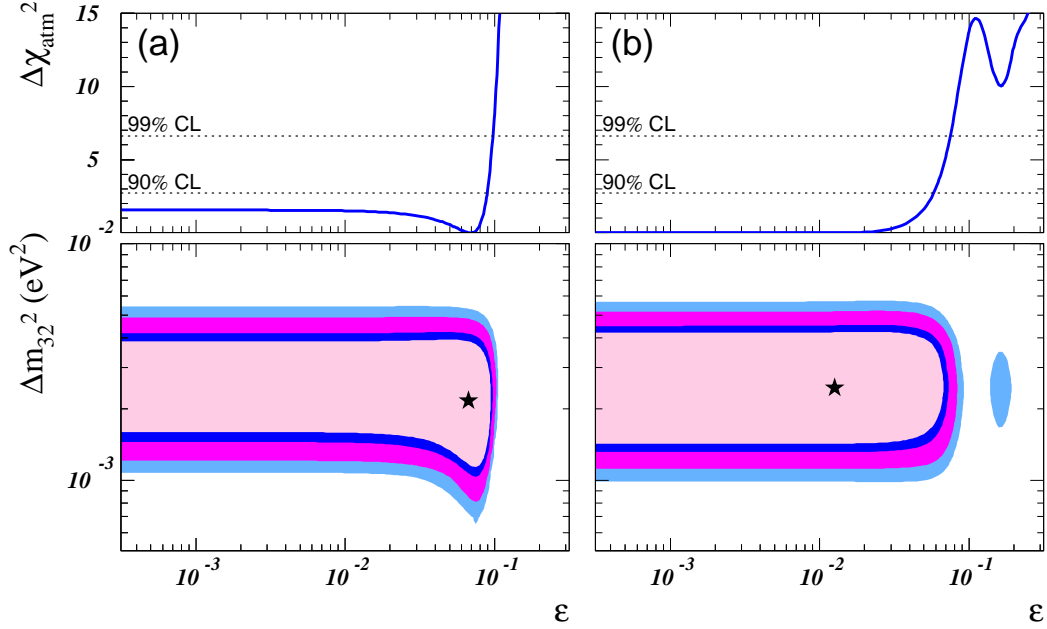}
    \caption{\label{fig:atm-epsi}%
      Allowed region (90\%, 95\%, 99\%, 99.73\% C.L.\ with 2 d.o.f.)
      in the plane $(\eps, \Delta m_{32}^2)$ from the analysis of
      atmospheric neutrino data, for each of the two {\it ans\"atze}
      given in Eqs.~\eqref{eq:ans_A} (left panels)
      and~\eqref{eq:ans_B} (right panels). The best fit point is
      denoted as a star. We fix the undisplayed parameters to
      $\tht_{13}=0$, $\tht_{23} = 45^\circ$ and $\eps' = 0.57$. The
      top panels show the behavior of $\Delta \chi_{\rm atm}^2$ as a
      function of $\eps$ when $\Delta m_{32}^2$ is integrated out.}
\end{figure}

In the lower panels of Fig.~\ref{fig:atm-epsi} we show the allowed
region in the plane $(\eps, \Delta m_{32}^2)$ from the analysis of
atmospheric neutrino data. The remaining undisplayed parameters are
set to $\tht_{13}=0$ (indicated by a combination of reactor and
atmospheric data~\cite{Apollonio:1999ae}), $\tht_{23} = 45^\circ$
(indicated by the standard atmospheric data analysis) and $\eps' =
0.57$ (indicated by the solar data analysis presented in
sections~\ref{sec:neutr-evol-pres} and \ref{sec:analysis-method}).
In the upper panels we show the dependence of $\Delta\chi_{\rm atm}^2$
as a function of $\eps$ when $\Delta m_{32}^2$ is ``integrated out''.

From Fig.~\ref{fig:atm-epsi} we see that in both cases the value of
$\Delta m_{32}^2$ is essentially unaffected by the inclusion of
non-standard interaction in the $\nu_\mu \to \nu_e$ and $\nu_\tau \to
\nu_e$ channels, up to values of 10 \% or so. This is analogous to
what was already found in Ref.~\cite{Val} for the $\nu_\mu \to
\nu_\tau$ channel.
We also note that in model $(a)$ values of $\eps \approx 0.07$ are
preferred, since in this case a small contribution of FC actually
helps in improving the SK contained $\nu_e$ data. This phenomenon is
analogous to having a small but finite solar mass splitting $\Delta
m_{21}^2$, see Ref.~\cite{Peres:1999yi}.
However, a value of $\eps=0$ is clearly in agreement with the data, as
expected by the excellent quality of the oscillation fit for the pure
$\nu_\mu \to \nu_\tau$ vacuum oscillations.
We see that in both schemes the values of $\eps$ allowed by the solar
data analysis are fully acceptable also for the atmospheric data.
However in model $(a)$ there is a small tension between the solar best
fit point ($\eps=0.003$) and the atmospheric data (which prefer, with
$\Delta\chi^2=1.6$, a value $\eps=0.07$), so that some slight change
in the shape of the solar allowed region might be expected when
atmospheric data are also included. This effect does not appear in
ansatz $(b)$, for which any value of $\eps$ below $0.03$ is
practically equivalent.

\section{Conclusions}
\label{sec:conclusions}

In this paper we have re-analysed the non-standard interaction
solutions to the solar neutrino problem in the light of the recent SNO
measurement as well as the 1258--day Super-Kamiokande data. In
contrast to previous work we have used a three-neutrino interpretation
of the solar data analysis, displaying how results depend on the
neutrino mixing angle $\tht_{13}$ which is constrained mainly by the
reactor neutrino data. More importantly, we have checked consistency
of the NSI interpretation of solar data with the good agreement that
the atmospheric data sample shows with the OSC interpretation.
We have found that the status of such energy-independent solution to
the solar neutrino problem is slightly better than that of the OSC
solution.  While these NSI solutions exist for reasonable values of
the flavour-changing interaction strength $ \eps_\eff$ they require a
somewhat large value of the non-universal parameters, suggesting that
the solar conversion channel must involve $\nu_\tau$. We have also
analysed the implications of the solar NSI solution for atmospheric
neutrinos, generalizing the study in Ref.~\cite{Val} so as to analyse
the sensitivity of atmospheric data also to the NSI parameters
involved in the solar neutrino channel. By using two simplified {\it
  ans\"atze} for the NSI interactions we have explicitly demonstrated
that the values of $\eps$ allowed by the solar data analysis are fully
acceptable also for the atmospheric data. This establishes that the
NSI description of the solar neutrino data is not in conflict with the
atmospheric data sample.
For one of these models we noted the existence of a slight conflict
between the solar best fit point ($\eps=0.003$) and the atmospheric
data (which prefer, with $\Delta\chi^2=1.6$, a value $\eps=0.07$),
which suggests that some slight change in the shape of the solar
allowed region might be expected within a fully global description of
both solar the atmospheric data.
Our simplified {\it ans\"atze} are justified to the extent that the
complexity of the analysis makes a full description close to
impossible, and possibly un-illuminating. We find it significant,
however, that the present understanding of solar and atmospheric
neutrino data does not, as yet, require solar neutrino oscillations or
solar neutrino mixing. This may have profound implications for
model-building, especially in view of the difficulties in accomodating
bi-large mixing-type solutions within the framework of unified
theories.  From an experimental point of view we find it worth
pointing that the energy dependence characteristic of the oscillation
mechanism has only been demonstrated for the case atmospheric, not
solar neutrino conversions, as yet.
One may argue that our NSI hypothesis is somewhat artificial. However
one should bear in mind the fact that most models neutrino masses are
accompanied by non-standard interactions of neutrinos and there are,
in fact, some in which NSI are unacompanied by neutrino mass effects.
The soundness of hybrid schemes such as the ones indicated here may
indicate more subtle new ways for accounting for the presently
observed neutrino anomalies from first principles.  We look forward to
new solar neutrino experiments to probe neutrino NSI with improved
sensitivities, such as suggested in Ref.~\cite{Berezhiani:2001rt}, as
well as to the prospects of future neutrino factories performing
correlated OSC-NSI studies, as suggested
in~\cite{Huber:2001de,Huber:2001zw}.

\section*{Acknowledgements}

This work was supported by Spanish DGICYT under grant PB98-0693
(Spain), FAPESP and CNPq (Brazil), by the European Commission RTN
network HPRN-CT-2000-00148 and by the European Science Foundation
network grant N.~86. M.M.\ was supported by the Marie Curie contract
HPMF-CT-2000-01008. M.A.T.\ was supported by the M.E.C.D.\ fellowship
AP2000-1953. We thank O. Peres, for useful discussions.


\section* {Appendix: Implications of the KamLAND result.}

The original version of the present paper showed that a pure NSI
description of solar data with massless and unmixed neutrinos is
slightly better than the favoured LMA-MSW solution. Moreover, the NSI
values indicated by the solar data analysis do not spoil the succesful
oscillation description of the atmospheric data, establishing the
overall consistency of a hybrid scheme in which only atmospheric data
are explained in terms of neutrino oscillations. However, the recent
results of the KamLAND collaboration~\cite{Eguchi:2002dm} reject
non-oscillation solutions in the solar sector under the assumption of
CPT invariance, in such a way that solutions based on NSI should be
strongly disfavoured by KamLAND results. In this appendix, we
re-evaluate the status of NSI solutions of the solar neutrino problem,
in the light of the first KamLAND data.

The KamLAND experiment is a reactor neutrino experiment whose detector
is located at the Kamiokande site. 
Most of the $\overline{\nu}_e$
flux incident at KamLAND comes from nuclear plants at distances 80-350
km from the detector, making the average baseline of about 180 km,
long enough to provide a sensitive probe of the LMA-MSW region. The
target for the $\overline{\nu}_e$ flux consists of a spherical
transparent balloon filled with 1000 tons of non-doped liquid
scintillator, and the antineutrinos are detected via the inverse
neutron $\beta$-decay process $\overline{\nu}_e + p \to e^+ + n$. The
KamLAND collaboration has for the first time measured the
disappearance of neutrinos produced in a power reactor. They observe a
strong evidence for the disappearance of neutrinos during their flight
over such distances, giving the first terrestrial confirmation of the
solar neutrino anomaly and also establishing the oscillation
hypothesis with man-produced neutrinos.


In this appendix we perform a combined analysis of the solar neutrino
data with the first KamLAND results in terms of neutrino non-standard
interactions with matter, setting for definiteness $\tht_{13} = 0$.
The analysis of solar neutrino data includes the most recent results
from all experiments~\cite{new-sol-data}, as well as the latest
measurements from SNO~\cite{new-sno} presented in the form of 34 data
bins (See Ref.~\cite{Maltoni:2002ni} for a detailed explanation of the
new solar analysis).  The details of the theoretical Monte-Carlo and
statistical analysis of the KamLAND results are given in
Ref.~\cite{Maltoni:2002aw}; in particular, the KamLAND
$\chi^2$-function is calculated assuming a Poisson distribution of the
experimental data, as described in Sec. IV of that paper.

In Table~\ref{tab:chi2-KL} we present the best-fit points for the
oscillation and NSI solutions to the solar neutrino problem obtained
in our analysis, together with its corresponding value of $\chi^2$.
The value of $\chi^2_\mathrm{sol}$ confirms our previous result: NSI
picture gives a good description of the solar neutrino data. In fact,
for the case of neutrino NSI with {\it up} quarks we obtain a slightly
better fit than for the LMA-MSW solution. 

The inclusion of the latest solar data
sample~\cite{new-sol-data,new-sno} does not change the status nor the
allowed regions of the NSI solution in any significant way.
In contrast, the inclusion of KamLAND data in the analysis changes
dramatically the situation.  Note that for the KamLAND experiment
matter effects are very small, and can therefore safely be neglected.
As a result, the NSI solutions predict no reduction in the
$\overline{\nu}_e$ flux, in conflict with what is observed at KamLAND.
This worsens the status of the NSI hypothesis, (reflected in the
corresponding $\chi^2_\mathrm{sol+KL}$ values, in
Table~\ref{tab:chi2-KL}) with respect to that of the LMA-MSW solution,
which predicts the correct suppression factor observed at KamLAND.
From our analysis we obtain that the description in terms of neutrino
NSI with {\it down} quarks is rejected with a $\Delta\chi^2$ = 14.2,
corresponding to 3.8$\sigma$, with respect to LMA-MSW, while NSI of
neutrinos with quarks of type {\it up} are rejected at 3.6$\sigma$
level ($\Delta\chi^2$ = 12.8) as the explanation for the solar
neutrino anomaly plus the KamLAND disappearance of neutrinos.

In summary, we have shown that the inclusion of KamLAND data in the
analysis of the solar neutrino anomaly excludes the interpretation
based on neutrino non-standard interactions with matter at more than
3$\sigma$. Therefore, non-standard interactions may at best play a
sub-leading role in solar neutrino propagation. In fact, one may use
the confirmation of the LMA-MSW oscillation solution together with the
experimental data from KamLAND and solar neutrino experiments in order
to determine improved restrictions on NSI parameters.

\begin{table}[t] \centering
    \catcode`?=\active \def?{\hphantom{0}}
    \newcommand{\E}[2]{{#1}\times 10^{#2}}
    \begin{tabular}{lcccccc}
        \hline
        \noalign{\smallskip}
        Solution & $\Delta m_{21}^2$ & $\tan^2(\tht_{12})$ &
        $\eps_\eff$ & $\eps'_\eff$ &
        $\chi^2_\mathrm{sol}$ & $\chi^2_\mathrm{sol + KL}$\\
        \noalign{\smallskip}
        \hline
        \noalign{\smallskip}
        LMA-MSW & $\E{7.2}{-5}$ & $0.46$  & $0$           & $0$     & $65.8$ & $71.9$ \\
        NSI (d) & $0$       & $0$     & $\E{3.2}{-3}$ & $0.62$  & $66.9$ & $86.1$ \\
        NSI (u) & $0$       & $0$     & $\E{1.3}{-3}$ & $0.44$  & $65.5$ & $84.7$ \\
        \noalign{\smallskip}
        \hline
    \end{tabular}
    \caption{\label{tab:chi2-KL}%
      Best-fit points of oscillation and NSI
      solutions to the solar neutrino problem before and after the
      inclusion of the KamLAND data in the analysis}
\end{table}
%


\end{document}